\documentclass[10pt,twocolumn]{article}
\usepackage{iccv}
\usepackage{times}
\usepackage{epsfig}
\usepackage{graphicx}
\usepackage{amsmath}
\usepackage{amssymb}
\usepackage{multirow}
\usepackage{tablefootnote}
\usepackage{threeparttable}
\usepackage[shortcuts]{extdash}
\usepackage{url}

\iccvfinalcopy 

\def\iccvPaperID{****} 

\begin{document}

\title{Loss Ensembles for Extremely Imbalanced Segmentation}
\author{Jun Ma\\
\\
Department of Mathematics, Nanjing University of Science and Technology, China
}



\maketitle

\begin{abstract}
This short paper briefly presents our methodology details of automatic intracranial aneurysms segmentation from brain MR scans.
We use ensembles of multiple models trained from different loss functions. Our method ranked first place in the ADAM challenge segmentation task.
The code and trained models are publicly available at
\url{https://github.com/JunMa11/ADAM2020}.
\end{abstract}

\section{Introduction: Task and Dataset}
Early detection of intracranial aneurysms, as well as accurate measurement and assessment of shape, plays a critical role in clinical routine, which can allow to monitor of the growth and rupture risk of aneurysms and make proper treatment decisions to be made~\cite{ADAM-Background}.
In the MICCAI 2020, ADAM (Aneurysm Detection And segMenation, \url{http://adam.isi.uu.nl/}) challenge was held to compare different methods of intracranial aneurysms detection and segmentation from Time of Flight Magnetic Resonance Angiographs (TOF-MRAs).
The purpose of this short paper is to present our method details for the intracranial aneurysms segmentation task.

The challenge organizers provide 113 cases for training and 141 cases for testing.
Each case includes one Time of Flight MRA (TOF-MRA) and one structural MR image (either T1, T2, or FLAIR). In the training set, 20 cases do not contain a diagnosed intracranial aneurysm. For the remaining 93 cases, manual annotations of unruptured intracranial aneurysms are provided in the form of binary masks. All manual annotations were made by experts in the field of medical annotations and checked by an experienced radiologist.
In the testing set, 117 cases have at least one untreated, unruptured intracranial aneurysm and 26 cases do not have intracranial aneurysms.
It should be noted that the testing set is not publicly released and participants are required to submit their method with a Docker container.

\section{Method}
\label{s:method}
The main difficulty in the aneurysms segmentation task is the extremely imbalanced problem.
Specifically, the median image size and aneurysm voxel size are $512\times512\times140$ and 238, respectively. The foreground ratio is $6.5\times10^{-6}$ where the aneurysm occupies very small partition in the whole image.

To handle the the extremely imbalanced problem, our key idea is to introduce `anti-imbalance' loss functions to training neural networks and then fuses different models with ensembles.
Our method is based on the well-known nnU-Net~\cite{isensee2020nnunet} and details are presented in the following subsections.

\subsection{Preprocessing}
We only use the official preprocessed TOF images to train 3D nnU-Net models. The preprocessing includes foreground (non-zero regions) cropping, rasampling, and Z-Score normalization, which are default settings in nnU-Net.

\subsection{Network architecture}
We use the default 3D full-resolution nnU-Net~\cite{isensee2020nnunet,cciccek2016UNet3D} and set the maximum channel number as 360 (default 320).

\subsection{Training protocols}
We train two groups five-fold cross-validation models by different loss function combinations~\cite{SegLossOdyssey}: Dice\cite{milletari2016vnet,diceV2} + Cross-Entropy (CE) loss, and Dice + TopK loss~\cite{wu2016topK}.
We train all the models on NVIDIA TITAN V100 GPU with a patch size of $256\times224\times56$ and a batch size of 2. Each fold costs about 5 days.

\subsection{Testing protocols}
Five best models during the cross-validation are select for ensembles.
To speed up the inference time, we disable the default test-time augmentation during inference.

\subsection{Post-processing}
We remove the predicted aneurysms less than 11 voxels from the segmentation results.

\section{Experimental Results}
\subsection{Cross-validation results}
Table~\ref{tab:results} presents the cross-validation results in terms of DSC scores. It can be found that none of the loss function can achieve the best DSC score among all five folds. Thus, we select the five best-fold models for final ensembles.

\begin{table}
\caption{Quantitative segmentation results (DSC) of different loss functions. The blod numbers denotes the best results.}\label{tab:results}
\centering
\begin{tabular}{lcc}
\hline
Fold & Dice + CE loss & DSC + TopK loss  \\ \hline
0    & 0.4370                    & \textbf{0.4921}                 \\
1    & \textbf{0.5476}                    & 0.4888                 \\
2    & \textbf{0.5108}                    & 0.4926                 \\
3    & \textbf{0.6173}                    & 0.5998                 \\
4    & 0.4404                    & \textbf{0.5240}                 \\
\hline
\end{tabular}
\end{table}

\subsection{Testing set results}
Table~\ref{tab:ADAM-test} shows the quantitative results of top-2 participant teams on ADAM Challenge Leaderboard \footnote{\url{http://adam.isi.uu.nl/results/results-miccai-2020/}}.
The team `junma' achieved the best DSC and Volumetric Similarity and the team `joker' achieved the best HD95. However, it should be noted that the differences between them are marginal.

\begin{table}[!htbp]
\caption{Quantitative results of top-2 participant teams on ADAM Challenge Leaderboard.}\label{tab:ADAM-test}
\centering
\begin{tabular}{lcccc}
\hline
Team  & DSC  & HD95 & Volumetric Similarity & Rank \\ \hline
junma & 0.41 & 8.96 & 0.50                   & 1    \\
joker & 0.40 & 8.67 & 0.48                   & 2    \\ \hline
\end{tabular}
\end{table}





\section{Conclusion}
This short paper briefly presents our solution to the aneurysms segmentation in the ADAM challenge. The main idea is to use ensembles of multiple models trained with different `anti-imbalance' loss functions. Our method ranked first place in the ADAM challenge segmentation task. However, the DSC score is still low, which has a large room for improvement.
In the future, we will extend this work and evaluate our method on more extremely imbalanced segmentation tasks.

\bibliographystyle{ieee}
\bibliography{egbib}

\begin{thebibliography}{1}\itemsep=-1pt

\bibitem{cciccek2016UNet3D}
{\"O}.~{\c{C}}i{\c{c}}ek, A.~Abdulkadir, S.~S. Lienkamp, T.~Brox, and
  O.~Ronneberger.
\newblock 3d u-net: learning dense volumetric segmentation from sparse
  annotation.
\newblock In {\em International Conference on Medical Image Computing and
  Computer-Assisted Intervention}, pages 424--432, 2016.

\bibitem{diceV2}
M.~Drozdzal, E.~Vorontsov, G.~Chartrand, S.~Kadoury, and C.~Pal.
\newblock The importance of skip connections in biomedical image segmentation.
\newblock In {\em Deep Learning and Data Labeling for Medical Applications},
  pages 179--187. Springer, 2016.

\bibitem{isensee2020nnunet}
F.~Isensee, P.~F. J{\"a}ger, S.~A. Kohl, J.~Petersen, and K.~H. Maier-Hein.
\newblock nnu-net: a self-configuring method for deep learning-based biomedical
  image segmentation.
\newblock {\em Nature Methods}, 2020.

\bibitem{SegLossOdyssey}
M.~Jun.
\newblock Segmentation loss odyssey.
\newblock {\em arXiv preprint arXiv:2005.13449}, 2020.

\bibitem{milletari2016vnet}
F.~Milletari, N.~Navab, and S.-A. Ahmadi.
\newblock V-net: Fully convolutional neural networks for volumetric medical
  image segmentation.
\newblock In {\em 2016 fourth international conference on 3D vision (3DV)},
  pages 565--571, 2016.

\bibitem{ADAM-Background}
J.~Wardlaw and P.~White.
\newblock The detection and management of unruptured intracranial aneurysms.
\newblock {\em Brain}, 123(2):205--221, 2000.

\bibitem{wu2016topK}
Z.~Wu, C.~Shen, and A.~v.~d. Hengel.
\newblock Bridging category-level and instance-level semantic image
  segmentation.
\newblock {\em arXiv preprint arXiv:1605.06885}, 2016.

\end{thebibliography}

\end{document}